\documentclass[zpreprint,zbstepj,british]{./zeus_Kpaper}
\usepackage{./zeusdet_units}
\usepackage{epsfig}
\chardef\usc=95
\chardef\til=126
\catcode`\@=11 
\DeclareRobustCommand\xdotspace{\futurelet\@let@token\@xdotspace}
\def\@xdotspace{%
  \ifx\@let@token.\else
  \ifx\@let@token\bgroup.\else
  \ifx\@let@token\egroup.\else
  \ifx\@let@token\/.\else
  \ifx\@let@token\ .\else
  \ifx\@let@token~.\else
  \ifx\@let@token!.\else
  \ifx\@let@token,.\else
  \ifx\@let@token:.\else
  \ifx\@let@token;.\else
  \ifx\@let@token?.\else
  \ifx\@let@token/.\else
  \ifx\@let@token'.\else
  \ifx\@let@token).\else
  \ifx\@let@token-.\else
  \ifx\@let@token\@xobeysp.\else
  \ifx\@let@token\space.\else
  \ifx\@let@token\@sptoken.\else
   .\space
   \fi\fi\fi\fi\fi\fi\fi\fi\fi\fi\fi\fi\fi\fi\fi\fi\fi\fi}
\catcode`\@=12 

\newcommand{\stru}[2]{%
   \relax\ifmmode\hbox{\vrule height#1 depth#2 width0pt}%
   \else\vrule height#1 depth#2 width0pt\fi}

\newcommand{\Ronum}[1]{\uppercase\expandafter{\romannumeral#1}}
\newcommand{\ronum}[1]{\expandafter{\romannumeral#1}}
\DeclareRobustCommand{\LaTeXZ}{%
  \LaTeX\kern-.05em4\kern-.1em
  {\raisebox{-0.2ex}{$\scriptstyle\text{ZEUS}$}}\xspace}

\newcommand{\slashfrac}[2]{%
  \raisebox{0.5ex}{\ensuremath #1}\kern-0.12em/\kern-0.08em
  \raisebox{-.8ex}{\ensuremath #2}}

\newcommand{\sqr}[3]{%
    {\vcenter{\hrule height.#3ex\hbox{\vrule width.#2ex height#1ex
     \kern#1ex\vrule width.#3ex}\hrule height.#2ex}}}

\catcode`\@=11 
\newcommand{\parenbar}{\mathpalette\p@renb@r}
\def\p@renb@r#1#2{\vbox{%
  \ifx#1\scriptscriptstyle \dimen@.7em\dimen@ii.2em\else
  \ifx#1\scriptstyle \dimen@.8em\dimen@ii.25em\else
  \dimen@1em\dimen@ii.4em\fi\fi \offinterlineskip
  \ialign{\hfill##\hfill\cr
    \vbox{\hrule width\dimen@ii}\cr
    \noalign{\vskip-.3ex}%
    \hbox to\dimen@{$\mathchar300\hfil\mathchar301$}\cr
    \noalign{\vskip-.3ex}%
    $#1#2$\cr}}}
\catcode`\@=12 

\ifzeusunit
  
\else
  
\fi

\newcommand{\IP}{{\rm I$\kern-0.01667em$P}\xspace}

\mathchardef\qsm=63
\mathchardef\pls=43
\mathchardef\mns=512
\mathchardef\plm=518
\mathchardef\eql=61
\mathchardef\smallleft=300
\mathchardef\smallright=301
\mathchardef\les=316
\mathchardef\gre=318
\mathchardef\leq=532
\mathchardef\grq=533

\catcode`\@=11 
\newcounter{pict@width}
\newcounter{pict@height}
\newlength{\pict@scale}
\newcommand{\psfigadd}[4]{%
\setcounter{pict@width}{1*\ratio{#2+\pict@scale/2}{\pict@scale}}
\setcounter{pict@height}{1*\ratio{#3+\pict@scale/2}{\pict@scale}}
\setlength{\unitlength}{\pict@scale}
\hbox to #2{\hspace{-\fill}\begin{picture}(\thepict@width,\thepict@height)
\put(0,0){\psfig{figure=#1,width=#2,height=#3,clip=}}
\SetScale{0.283466457}
\SetWidth{1.763889}
{#4}
\end{picture}}
}
\newcounter{pict@widthfst}
\newcounter{pict@widthscd}
\newcounter{pict@widthtot}
\newcommand{\psfigaddtwo}[7]{%
\setcounter{pict@widthfst}{1*\ratio{#2+\pict@scale/2}{\pict@scale}}
\setcounter{pict@widthscd}{1*\ratio{#2+#4+\pict@scale/2}{\pict@scale}}
\setcounter{pict@widthtot}{1*\ratio{#2+#4+#6+\pict@scale/2}{\pict@scale}}
\setcounter{pict@height}{1*\ratio{#3+\pict@scale/2}{\pict@scale}}
\setlength{\unitlength}{\pict@scale}
\hbox{\hspace{-\fill}\begin{picture}(\thepict@widthtot,\thepict@height)
\put(0,0){\psfig{figure=#1,width=#2,height=#3,clip=}}
\put(\thepict@widthscd,0){\psfig{figure=#5,width=#6,height=#3,clip=}}
\SetScale{0.283466457}
\SetWidth{1.763889}
{#7}
\end{picture}}
}
\newcommand{\psfigror}[4]{%
\setcounter{pict@width}{1*\ratio{#2+\pict@scale/2}{\pict@scale}}
\setcounter{pict@height}{1*\ratio{#3+\pict@scale/2}{\pict@scale}}
\setlength{\unitlength}{\pict@scale}
\hbox{\begin{picture}(\thepict@width,\thepict@height)
\put(0,\thepict@height){\psfig{figure=#1,width=#3,height=#2,clip=,angle=270}}
\SetScale{0.283466457}
\SetWidth{1.763889}
{#4}
\end{picture}}
}
\newcommand{\psfigrol}[4]{%
\setcounter{pict@width}{1*\ratio{#2+\pict@scale/2}{\pict@scale}}
\setcounter{pict@height}{1*\ratio{#3+\pict@scale/2}{\pict@scale}}
\setlength{\unitlength}{\pict@scale}
\hbox{\begin{picture}(\thepict@width,\thepict@height)
\put(0,0){\psfig{figure=#1,width=#3,height=#2,clip=,angle=90}}
\SetScale{0.283466457}
\SetWidth{1.763889}
{#4}
\end{picture}}
}
\catcode`\@=12 
\newlength\listtextwidth

\catcode`\@=11 
\newlength{\@tabfninsert}
\newlength{\@tabfnwidth}
\newcommand{\tabfootnote}[2]{%
  \setlength{\@tabfninsert}{0.8em}
  \setlength{\@tabfnwidth}{\textwidth}
  \addtolength{\@tabfnwidth}{-\@tabfninsert}
  \addtolength{\@tabfnwidth}{-0.4em}
  \noindent\makebox[\@tabfninsert][r]{\footnotesize$^{#1}$\hfil}\hfill%
  \parbox[t]{\@tabfnwidth}{\footnotesize #2\hfill}}
\catcode`\@=12 

\def\citeCTD{{\cite{%
nim:a279:290,*npps:b32:181,*nim:a338:254%
}}\xspace}
\def\citeMVD{{\cite{%
nim:a581:656%
}}\xspace}
\def\citeSTT{{\cite{%
nim:a535:191%
}}\xspace}
\def\citeCAL{{\cite{%
nim:a309:77,*nim:a309:101,*nim:a321:356,*nim:a336:23%
}}\xspace}
\def\citeHES{{\cite{%
nim:a277:176%
}}\xspace}
\def\citeSixmT{{\cite{%
thesis:gosau:2007%
}}\xspace}
\def\citeBAC{{\cite{%
nim:a300:480%
}}\xspace}
\def\citePCAL{{\cite{%
desy-92-066,*zfp:c63:391,*acpp:b32:2025%
}}\xspace}
\def\citeSPECTRO{{\cite{%
nim:a565:572%
}}\xspace}

\begin{document}
\prepnum{DESY--14--082}
\prepdate{September 2014}

\zeustitle{%
  Measurement of \boldmath{$D^*$} photoproduction at three different centre-of-mass 
  energies at HERA
}

\zeusauthor{ZEUS Collaboration}
\draftversion{2.0}
\zeusdate{}

\maketitle

\begin{abstract}\noindent
The photoproduction of $D^{*\pm}$ mesons has been measured 
with the ZEUS detector at HERA at three different $ep$ centre-of-mass energies, 
$\sqrt{s}$, of 318, 251 and 225\,GeV.  For each data set, $D^{*\pm}$ mesons were 
required to have transverse momentum, $p_T^{D^*}$, and pseudorapidity, $\eta^{D^*}$, 
in the ranges \mbox{$1.9 < p_T^{D^*} < 20$\,GeV} and $|\eta^{D^*}|<1.6$.  The 
events were required to have a virtuality of the incoming photon, $Q^2$, of less 
than 1\,GeV$^2$.  
The dependence on $\sqrt{s}$ was studied by normalising to the high-statistics measurement 
at $\sqrt{s} =318$\,GeV.  This led to the cancellation of a number of 
systematic effects both in data and theory.  Predictions from next-to-leading-order 
QCD describe the $\sqrt{s}$ dependence of the data well.
\end{abstract}

\thispagestyle{empty}
\cleardoublepage

%
%
%
%

\def\3{\ss}
\pagenumbering{Roman}
                                                   %
\begin{center}
{                      \Large  The ZEUS Collaboration              }
\end{center}

{\small


        {\raggedright
H.~Abramowicz$^{27, u}$, 
I.~Abt$^{21}$, 
L.~Adamczyk$^{8}$, 
M.~Adamus$^{34}$, 
R.~Aggarwal$^{4, a}$, 
S.~Antonelli$^{2}$, 
O.~Arslan$^{3}$, 
V.~Aushev$^{16, 17, o}$, 
Y.~Aushev$^{17, o, p}$, 
O.~Bachynska$^{10}$
% ...
        }

\newpage


\makebox[3em]{$^{1}$}
\begin{minipage}[t]{14cm}
{\it INFN Bologna, Bologna, Italy}~$^{A}$

\end{minipage}\\
\makebox[3em]{$^{2}$}
\begin{minipage}[t]{14cm}
{\it University and INFN Bologna, Bologna, Italy}~$^{A}$

\end{minipage}\\
\makebox[3em]{$^{3}$}
\begin{minipage}[t]{14cm}
{\it Physikalisches Institut der Universit\"at Bonn,
Bonn, Germany}~$^{B}$

\end{minipage}\\
\makebox[3em]{$^{4}$}
\begin{minipage}[t]{14cm}
{\it Panjab University, Department of Physics, Chandigarh, India}

\end{minipage}\\
\makebox[3em]{$^{5}$}
\begin{minipage}[t]{14cm}
{\it Calabria University,
Physics Department and INFN, Cosenza, Italy}~$^{A}$

\end{minipage}\\
\makebox[3em]{$^{6}$}
\begin{minipage}[t]{14cm}
{\it National Centre for Particle Physics, Universiti Malaya, 50603 Kuala Lumpur, Malaysia}~$^{C}$

\end{minipage}\\
\makebox[3em]{$^{7}$}
\begin{minipage}[t]{14cm}
{\it The Henryk Niewodniczanski Institute of Nuclear Physics, Polish Academy of \\
Sciences, Krakow, Poland}~$^{D}$

\end{minipage}\\
\makebox[3em]{$^{8}$}
\begin{minipage}[t]{14cm}
{\it AGH-University of Science and Technology, Faculty of Physics and Applied Computer
Science, Krakow, Poland}~$^{D}$

\end{minipage}\\
\makebox[3em]{$^{9}$}
\begin{minipage}[t]{14cm}
{\it Department of Physics, Jagellonian University, Cracow, Poland}

\end{minipage}\\
\makebox[3em]{$^{10}$}
\begin{minipage}[t]{14cm}
{\it Deutsches Elektronen-Synchrotron DESY, Hamburg, Germany}

\end{minipage}\\
\makebox[3em]{$^{11}$}
\begin{minipage}[t]{14cm}
{\it Deutsches Elektronen-Synchrotron DESY, Zeuthen, Germany}

\end{minipage}\\
\makebox[3em]{$^{12}$}
\begin{minipage}[t]{14cm}
{\it School of Physics and Astronomy, University of Glasgow,
Glasgow, United Kingdom}~$^{E}$

\end{minipage}\\
\makebox[3em]{$^{13}$}
\begin{minipage}[t]{14cm}
{\it Hamburg University, Institute of Experimental Physics, Hamburg,
Germany}~$^{F}$

\end{minipage}\\
\makebox[3em]{$^{14}$}
\begin{minipage}[t]{14cm}
{\it Institute of Particle and Nuclear Studies, KEK,
Tsukuba, Japan}~$^{G}$

\end{minipage}\\
\makebox[3em]{$^{15}$}
\begin{minipage}[t]{14cm}
{\it Institute of Physics and Technology of Ministry of Education and
Science of Kazakhstan, Almaty, Kazakhstan}

\end{minipage}\\
\makebox[3em]{$^{16}$}
\begin{minipage}[t]{14cm}
{\it Institute for Nuclear Research, National Academy of Sciences, Kyiv, Ukraine}

\end{minipage}\\
\makebox[3em]{$^{17}$}
\begin{minipage}[t]{14cm}
{\it Department of Nuclear Physics, National Taras Shevchenko University of Kyiv, Kyiv, Ukraine}

\end{minipage}\\
\makebox[3em]{$^{18}$}
\begin{minipage}[t]{14cm}
{\it Department of Physics, McGill University,
Montr\'eal, Qu\'ebec, Canada H3A 2T8}~$^{H}$

\end{minipage}\\
\makebox[3em]{$^{19}$}
\begin{minipage}[t]{14cm}
{\it Meiji Gakuin University, Faculty of General Education,
Yokohama, Japan}~$^{G}$

\end{minipage}\\
\makebox[3em]{$^{20}$}
\begin{minipage}[t]{14cm}
{\it Lomonosov Moscow State University, Skobeltsyn Institute of Nuclear Physics,
Moscow, Russia}~$^{I}$

\end{minipage}\\
\makebox[3em]{$^{21}$}
\begin{minipage}[t]{14cm}
{\it Max-Planck-Institut f\"ur Physik, M\"unchen, Germany}

\end{minipage}\\
\makebox[3em]{$^{22}$}
\begin{minipage}[t]{14cm}
{\it Department of Physics, University of Oxford,
Oxford, United Kingdom}~$^{E}$

\end{minipage}\\
\makebox[3em]{$^{23}$}
\begin{minipage}[t]{14cm}
{\it INFN Padova, Padova, Italy}~$^{A}$

\end{minipage}\\
\makebox[3em]{$^{24}$}
\begin{minipage}[t]{14cm}
{\it Dipartimento di Fisica dell' Universit\`a and INFN,
Padova, Italy}~$^{A}$

\end{minipage}\\
\makebox[3em]{$^{25}$}
\begin{minipage}[t]{14cm}
{\it Polytechnic University, Tokyo, Japan}~$^{G}$

\end{minipage}\\
\makebox[3em]{$^{26}$}
\begin{minipage}[t]{14cm}
{\it Dipartimento di Fisica, Universit\`a `La Sapienza' and INFN,
Rome, Italy}~$^{A}$

\end{minipage}\\
\makebox[3em]{$^{27}$}
\begin{minipage}[t]{14cm}
{\it Raymond and Beverly Sackler Faculty of Exact Sciences, School of Physics, \\
Tel Aviv University, Tel Aviv, Israel}~$^{J}$

\end{minipage}\\
\makebox[3em]{$^{28}$}
\begin{minipage}[t]{14cm}
{\it Department of Physics, Tokyo Institute of Technology,
Tokyo, Japan}~$^{G}$

\end{minipage}\\
\makebox[3em]{$^{29}$}
\begin{minipage}[t]{14cm}
{\it Universit\`a di Torino and INFN, Torino, Italy}~$^{A}$

\end{minipage}\\
\makebox[3em]{$^{30}$}
\begin{minipage}[t]{14cm}
{\it Universit\`a del Piemonte Orientale, Novara, and INFN, Torino,
Italy}~$^{A}$

\end{minipage}\\
\makebox[3em]{$^{31}$}
\begin{minipage}[t]{14cm}
{\it Department of Physics, University of Toronto, Toronto, Ontario,
Canada M5S 1A7}~$^{H}$

\end{minipage}\\
\makebox[3em]{$^{32}$}
\begin{minipage}[t]{14cm}
{\it Physics and Astronomy Department, University College London,
London, United Kingdom}~$^{E}$

\end{minipage}\\
\makebox[3em]{$^{33}$}
\begin{minipage}[t]{14cm}
{\it Faculty of Physics, University of Warsaw, Warsaw, Poland}

\end{minipage}\\
\makebox[3em]{$^{34}$}
\begin{minipage}[t]{14cm}
{\it National Centre for Nuclear Research, Warsaw, Poland}

\end{minipage}\\
\makebox[3em]{$^{35}$}
\begin{minipage}[t]{14cm}
{\it Department of Particle Physics and Astrophysics, Weizmann
Institute, Rehovot, Israel}

\end{minipage}\\
\makebox[3em]{$^{36}$}
\begin{minipage}[t]{14cm}
{\it Department of Physics, York University, Ontario, Canada M3J 1P3}~$^{H}$

\end{minipage}\\


\makebox[3ex]{$^{ A}$}
\begin{minipage}[t]{14cm}
 supported by the Italian National Institute for Nuclear Physics (INFN) \
\end{minipage}\\
\makebox[3ex]{$^{ B}$}
\begin{minipage}[t]{14cm}
 supported by the German Federal Ministry for Education and Research (BMBF), under
 contract No. 05 H09PDF\
\end{minipage}\\
\makebox[3ex]{$^{ C}$}
\begin{minipage}[t]{14cm}
 supported by HIR grant UM.C/625/1/HIR/149 and UMRG grants RU006-2013, RP012A-13AFR and RP012B-13AFR from
 Universiti Malaya, and ERGS grant ER004-2012A from the Ministry of Education, Malaysia\
\end{minipage}\\
\makebox[3ex]{$^{ D}$}
\begin{minipage}[t]{14cm}
 supported by the National Science Centre under contract No. DEC-2012/06/M/ST2/00428\
\end{minipage}\\
\makebox[3ex]{$^{ E}$}
\begin{minipage}[t]{14cm}
 supported by the Science and Technology Facilities Council, UK\
\end{minipage}\\
\makebox[3ex]{$^{ F}$}
\begin{minipage}[t]{14cm}
 supported by the German Federal Ministry for Education and Research (BMBF), under
 contract No. 05h09GUF, and the SFB 676 of the Deutsche Forschungsgemeinschaft (DFG) \
\end{minipage}\\
\makebox[3ex]{$^{ G}$}
\begin{minipage}[t]{14cm}
 supported by the Japanese Ministry of Education, Culture, Sports, Science and Technology
 (MEXT) and its grants for Scientific Research\
\end{minipage}\\
\makebox[3ex]{$^{ H}$}
\begin{minipage}[t]{14cm}
 supported by the Natural Sciences and Engineering Research Council of Canada (NSERC) \
\end{minipage}\\
\makebox[3ex]{$^{ I}$}
\begin{minipage}[t]{14cm}
 supported by RF Presidential grant N 3042.2014.2 for the Leading Scientific Schools and by
 the Russian Ministry of Education and Science through its grant for Scientific Research on
 High Energy Physics\
\end{minipage}\\
\makebox[3ex]{$^{ J}$}
\begin{minipage}[t]{14cm}
 supported by the Israel Science Foundation\
\end{minipage}\\
\vspace{30em} \pagebreak[4]


\makebox[3ex]{$^{ a}$}
\begin{minipage}[t]{14cm}
also funded by Max Planck Institute for Physics, Munich, Germany\
\end{minipage}\\
\makebox[3ex]{$^{ b}$}
\begin{minipage}[t]{14cm}
also funded by Max Planck Institute for Physics, Munich, Germany, now at Sri Guru Granth Sahib World University, Fatehgarh Sahib\
\end{minipage}\\
\makebox[3ex]{$^{ c}$}
\begin{minipage}[t]{14cm}
also at Agensi Nuklear Malaysia, 43000 Kajang, Bangi, Malaysia\
\end{minipage}\\
\makebox[3ex]{$^{ d}$}
\begin{minipage}[t]{14cm}
also at Cracow University of Technology, Faculty of Physics, Mathematics and Applied Computer Science, Poland\
\end{minipage}\\
\makebox[3ex]{$^{ e}$}
\begin{minipage}[t]{14cm}
partially supported by the Polish National Science Centre projects DEC-2011/01/B/ST2/03643 and DEC-2011/03/B/ST2/00220\
\end{minipage}\\
\makebox[3ex]{$^{ f}$}
\begin{minipage}[t]{14cm}
now at Rockefeller University, New York, NY 10065, USA\
\end{minipage}\\
\makebox[3ex]{$^{ g}$}
\begin{minipage}[t]{14cm}
now at University of Liverpool, United Kingdom\
\end{minipage}\\
\makebox[3ex]{$^{ h}$}
\begin{minipage}[t]{14cm}
also affiliated with University College London, UK\
\end{minipage}\\
\makebox[3ex]{$^{ i}$}
\begin{minipage}[t]{14cm}
supported by the Alexander von Humboldt Foundation\
\end{minipage}\\
\makebox[3ex]{$^{ j}$}
\begin{minipage}[t]{14cm}
now at CERN, Geneva, Switzerland\
\end{minipage}\\
\makebox[3ex]{$^{ k}$}
\begin{minipage}[t]{14cm}
also at Institute of Theoretical and Experimental Physics, Moscow, Russia\
\end{minipage}\\
\makebox[3ex]{$^{ l}$}
\begin{minipage}[t]{14cm}
Alexander von Humboldt Professor; also at DESY and University of Oxford\
\end{minipage}\\
\makebox[3ex]{$^{ m}$}
\begin{minipage}[t]{14cm}
also at University of Tokyo, Japan\
\end{minipage}\\
\makebox[3ex]{$^{ n}$}
\begin{minipage}[t]{14cm}
now at Kobe University, Japan\
\end{minipage}\\
\makebox[3ex]{$^{ o}$}
\begin{minipage}[t]{14cm}
supported by DESY, Germany\
\end{minipage}\\
\makebox[3ex]{$^{ p}$}
\begin{minipage}[t]{14cm}
member of National Technical University of Ukraine, Kyiv Polytechnic Institute, Kyiv, Ukraine\
\end{minipage}\\
\makebox[3ex]{$^{ q}$}
\begin{minipage}[t]{14cm}
now at DESY ATLAS group\
\end{minipage}\\
\makebox[3ex]{$^{ r}$}
\begin{minipage}[t]{14cm}
member of National University of Kyiv - Mohyla Academy, Kyiv, Ukraine\
\end{minipage}\\
\makebox[3ex]{$^{ s}$}
\begin{minipage}[t]{14cm}
now at BNL, USA\
\end{minipage}\\
\makebox[3ex]{$^{ t}$}
\begin{minipage}[t]{14cm}
now at LNF, Frascati, Italy\
\end{minipage}\\
\makebox[3ex]{$^{ u}$}
\begin{minipage}[t]{14cm}
also at Max Planck Institute for Physics, Munich, Germany, External Scientific Member\
\end{minipage}\\
\makebox[3ex]{$^{ v}$}
\begin{minipage}[t]{14cm}
now at Tokyo Metropolitan University, Japan\
\end{minipage}\\
\makebox[3ex]{$^{ w}$}
\begin{minipage}[t]{14cm}
also supported by DESY\
\end{minipage}\\
\makebox[3ex]{$^{ x}$}
\begin{minipage}[t]{14cm}
also at \L\'{o}d\'{z} University, Poland\
\end{minipage}\\
\makebox[3ex]{$^{ y}$}
\begin{minipage}[t]{14cm}
member of \L\'{o}d\'{z} University, Poland\
\end{minipage}\\
\makebox[3ex]{$^{ z}$}
\begin{minipage}[t]{14cm}
now at Polish Air Force Academy in Deblin\
\end{minipage}\\
\makebox[3ex]{$^{aa}$}
\begin{minipage}[t]{14cm}
deceased\
\end{minipage}\\
\pagebreak[4]
}

%
\pagenumbering{arabic}
%
%
\section{Introduction}
\label{sec-int}

The photoproduction of charm quarks at HERA is a rich testing ground for the 
predictions of perturbative quantum chromodynamics (pQCD).  The predictions are 
expected to be reliable since the charm mass provides a hard scale in the 
perturbative expansion.  The dominant production mechanism is boson--gluon 
fusion.  Many measurements of charm photoproduction at high $ep$ centre-of-mass 
energies, $\sqrt{s}=318$\,GeV or $\sqrt{s}=300$\,GeV, have been made at 
HERA~\cite{epj:c72:2047,*epj:c72:1995,*epj:c50:251,*epj:c47:597,*pl:b621:56,*np:b472:32,*np:b866:229,*epj:c60:25,*np:b729:492,*pl:b565:87,*pl:b481:213,*pl:b401:192,*pl:b349:225,epj:c6:67,epj:c71:1659,pr:d78:072001} and compared with QCD predictions at next-to-leading order (NLO).  The 
description of the data is generally reasonable, although the uncertainties on 
the theory are often large.   

Previous results on charm photoproduction were obtained at a single $ep$ 
centre-of-mass energy; the dependence on the $ep$ centre-of-mass energy is presented 
here for the first 
time.  The variation of the cross section with centre-of-mass energy is sensitive 
to the gluon distribution in the proton, as different values of Bjorken $x$ are probed.  
Measurements of $D^{*\pm}$ production at three different centre-of-mass energies,
$\sqrt{s} =318$, 251 and 225\,GeV, are presented in this paper.  The variation of $\sqrt{s}$ 
was achieved by varying the proton beam energy, $E_p$, while keeping the 
electron\footnote{Hereafter ``electron'' refers to both electrons and positrons 
unless otherwise stated.} beam energy constant, $E_e = 27.5$\,GeV.  The data were collected 
in 2006 and 2007 with $E_p =920$, 575 and 460\,GeV, referred to, respectively, as the 
high- (HER), medium- (MER) and low-energy-running (LER) samples.  The corresponding 
luminosities of the HER, MER and LER samples are 144, 6.3 and 13.4\,pb$^{-1}$, 
respectively.  A common analysis procedure is used for all 
samples and the cross sections at different $\sqrt{s}$ are presented normalised to 
that for the HER data, thereby leading to a cancellation of a number of systematic uncertainties 
both in data and theory.

\section{Experimental set-up}
\label{sec-exp}

\Zdetdesc

\Zctdmvddesc{\ZcoosysfnBEeta}

\Zcaldesc

\ZlumidescA{1.8}


\section{Event selection and signal extraction}
\label{sec:selection}

\subsection{Photoproduction event selection}
\label{sec:event-selection}

A three-level trigger system~\cite{zeus:1993:bluebook,nim:a580:1257,uproc:chep:1992:222} 
was used to select events online.  The first- and second-level trigger used CAL and CTD 
data to select $ep$ collisions and to reject beam-gas events.  At the third level, the 
full event information was available.  In this analysis, triggers containing a $D$-meson 
candidate and/or two jets were used.

In order to remove non-$ep$ background, the $Z$ position of the primary vertex of an event, 
$Z_{\rm vtx}$, was required to be in the range $|Z_{\rm vtx}| < 30$\,cm.
Photoproduction events were selected by requiring that no scattered electron with energy 
larger than 5\,GeV was found in the CAL~\cite{pl:b322:287}.

The fraction of the incoming electron momentum carried by the photon, $y$, was 
reconstructed via the Jacquet--Blondel~\cite{proc:epfacility:1979:391} estimator, 
$y_{\rm JB}$, using energy-flow objects (EFOs)~\cite{epj:c6:43,*thesis:briskin:1998}. 
Energy-flow objects combine track and calorimeter information to optimise the resolution 
of the variable.  The value of $y_{\rm JB}$ is given by 
$y_{\rm JB} = \sum_i E_i (1 - \cos\theta_i)/2E_e$ where $E_e$ is the energy of the electron 
beam, $E_i$ is the energy of the $i$-th EFO, $\theta_i$ is its polar angle and the sum runs 
over all EFOs. The range 
$0.167 < y_{\rm JB} < 0.802$ was used, where the lower cut was set by the trigger 
requirements and the upper cut suppressed remaining events from deep inelastic scattering  
with an unidentified low-energy scattered electron in the CAL.  The range in $y_{\rm JB}$ 
corresponds to reconstructed photon--proton centre-of-mass energy, $W_{\rm JB}$, ranges of 
$130 < W_{\rm JB} < 285$\,GeV, $103 < W_{\rm JB} < 225$\,GeV and 
$92 < W_{\rm JB} < 201$\,GeV for the HER, MER, LER samples, respectively.

\subsection{Selection of \boldmath{$D^{*\pm}$} candidates and signal extraction}
\label{sec:dstar-selection}

The $D^{*+}$ mesons\footnote{Hereafter the charge conjugated states are implied.} were 
identified using the decay channel $D^{*+} \to D^0 \pi_s^+$ with the subsequent decay 
$D^0 \to K^- \pi^+$, where $\pi_s^+$ refers to a low-momentum (``slow'') pion 
accompanying the $D^0$.  Tracks from the $D^{*+}$ decay products were required to have 
at least one hit in the 
MVD and in the inner superlayer of the CTD and to reach at least the third CTD superlayer.  
Tracks with opposite charge and with  transverse momentum $p_T^{K,\pi}>0.4$\,GeV were 
combined in pairs to form $D^0$ candidates.  The tracks were alternately assigned the 
kaon and pion mass and the invariant mass of the pair, $M(K\pi)$, was calculated. Each 
additional track with charge opposite to that of the kaon track and a transverse 
momentum $p_T^{\pi_s}>0.12$\,GeV was assigned the pion mass and combined with the $D^0$ 
candidate to form a $D^{*+}$ candidate.  
Since more combinatorial background exists in the forward direction as well as 
at low $p_T^{D^*}$~\cite{epj:c6:67}, this was suppressed by 
requiring $p_T^{D^{*}}/E_T^{\theta>10^\circ} > 0.12$, where $p_T^{D^*}$ is the transverse 
momentum of the $D^{*+}$ meson and $E_T^{\theta>10^\circ}$ is the 
transverse energy measured using all CAL cells outside a cone of $10^\circ$ around the 
forward direction.  The mass difference 
$\Delta M \equiv M(K\pi\pi_s) - M(K\pi)$ was used to extract the $D^{*+}$ signal.  The 
$D^{*+}$ candidates were required to have \mbox{$1.83 < M(K\pi) <1.90$\,GeV}, 
$0.143<\Delta M<0.148$\,GeV, $1.9 < p_T^{D^*}<20$\,GeV and pseudorapidity, 
\mbox{$|\eta^{ D^*}|<1.6$}.  To allow the background to 
be determined, $D^0$ candidates with wrong-sign combinations, in which both 
tracks forming the $D^0$ candidates have the same charge and the third track has 
the opposite charge, were also retained. The same kinematic restrictions were 
applied as for those $D^0$ candidates with correct-charge combinations.

\begin{figure}[p]
\begin{center}
\epsfig{file=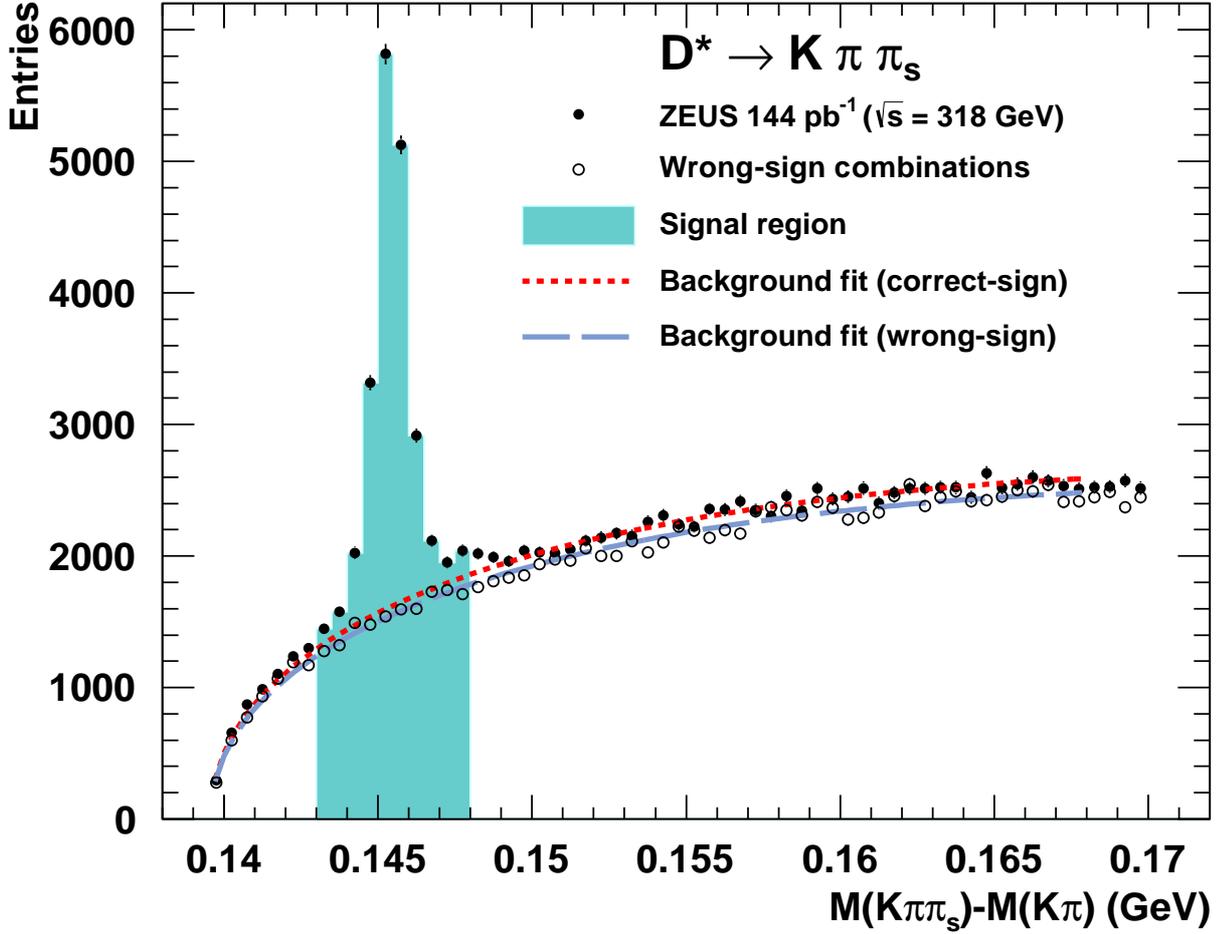,width=\textwidth}
\end{center}
\caption{
Distribution of the mass difference, $\Delta M \equiv M(K\pi\pi_s) - M(K\pi)$, 
for the $D^{* \pm}$ candidates for the HER ($\sqrt{s}=318$\,GeV) data sample.  The candidates 
are shown for correct-sign (filled circles) and wrong-sign 
combinations (empty circles).  The background fit is shown as a short-dashed (long-dashed) 
line for correct-sign (wrong-sign) combinations.  The $D^{*\pm}$ signal region is marked 
as a shaded area.}
\label{fig:dstar-her}
\end{figure}

\begin{figure}[p]
\begin{center}
\epsfig{file=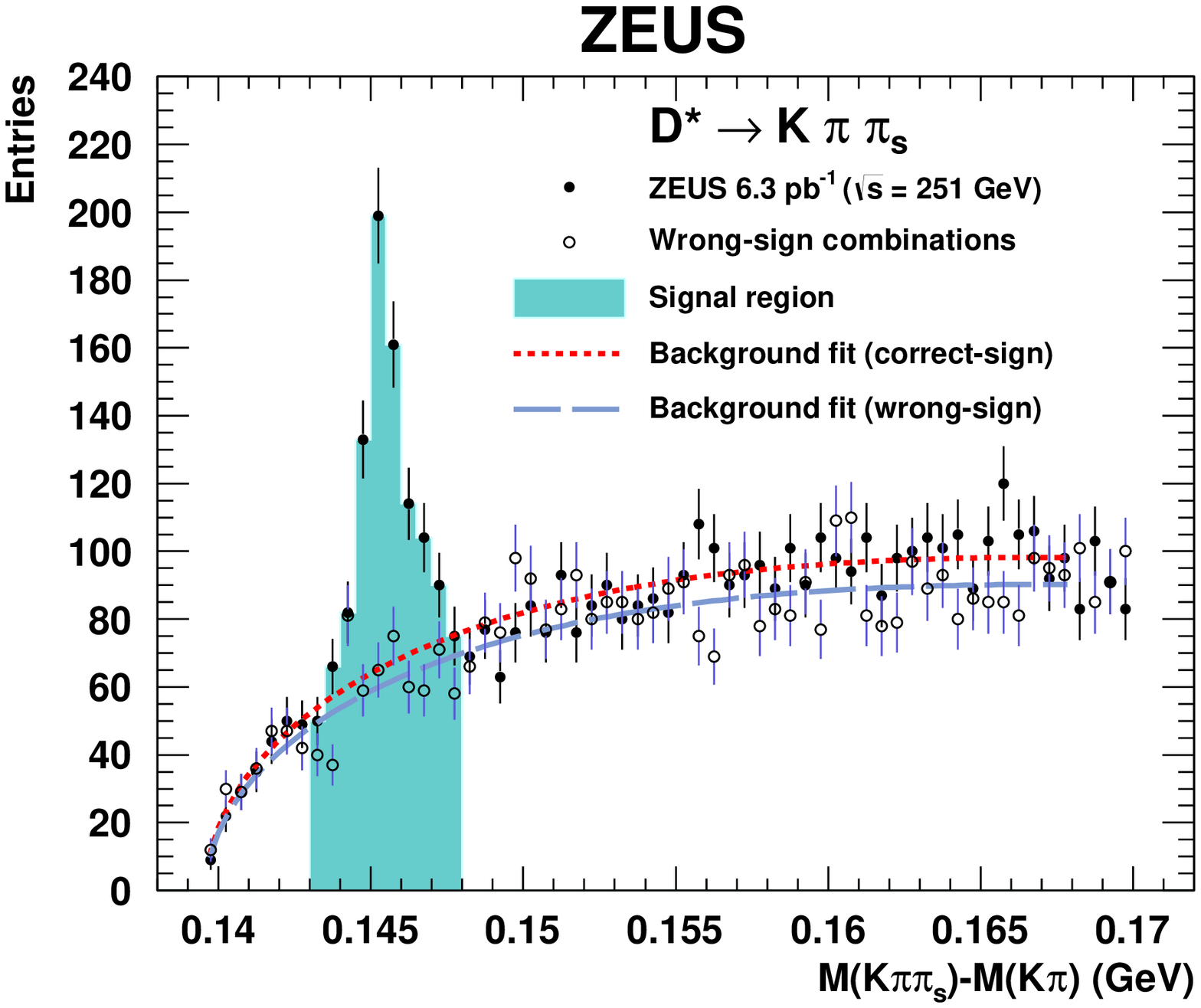,width=\textwidth}
\end{center}
\caption{
Distribution of the mass difference, $\Delta M \equiv M(K\pi\pi_s) - M(K\pi)$, 
for the $D^{* \pm}$ candidates for the MER ($\sqrt{s}=251$\,GeV) data sample.  Other 
details as in Fig.~\ref{fig:dstar-her}.}
\label{fig:dstar-mer}
\end{figure}

\begin{figure}[p]
\begin{center}
\epsfig{file=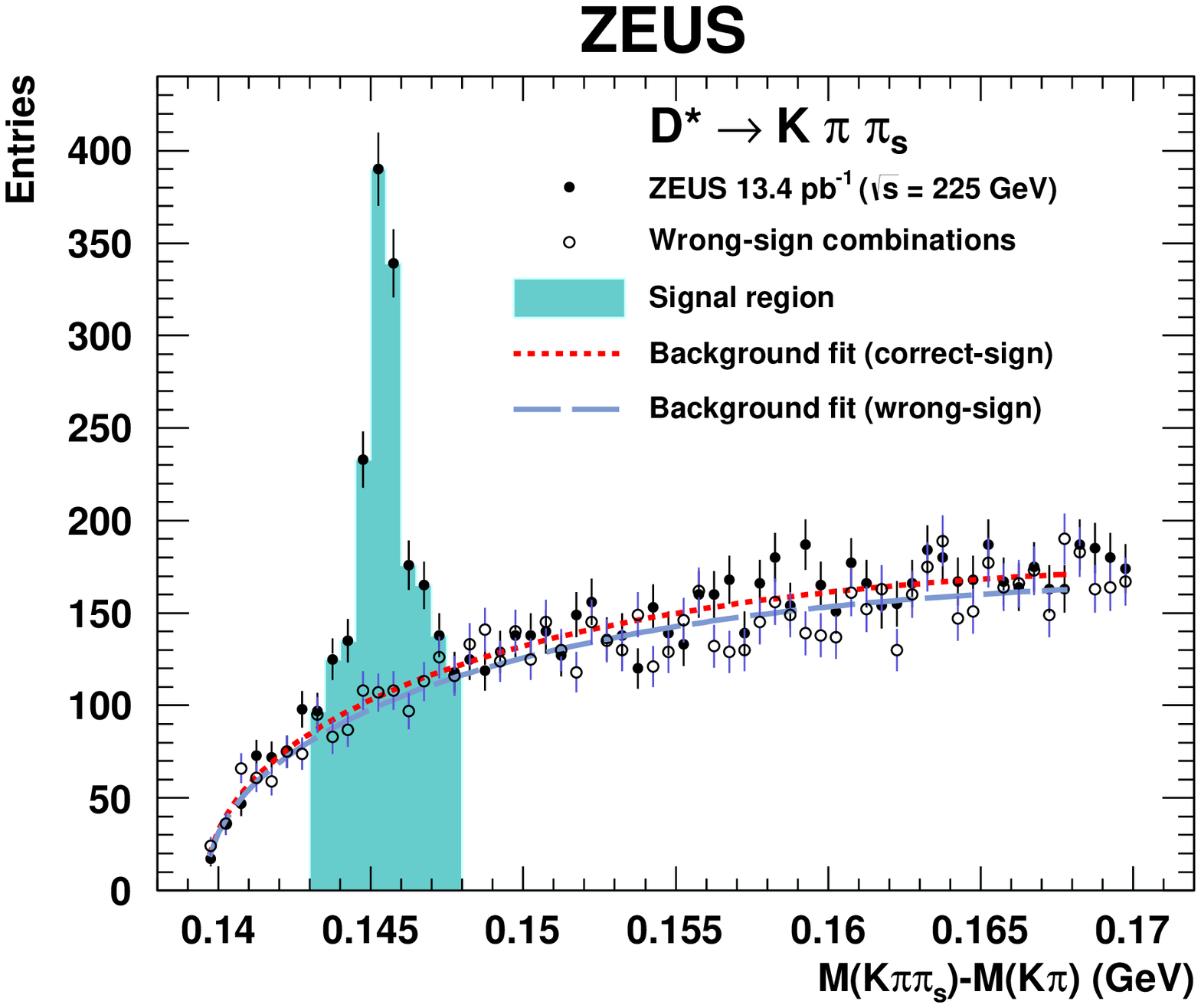,width=\textwidth}
\end{center}
\caption{
Distribution of the mass difference, $\Delta M \equiv M(K\pi\pi_s) - M(K\pi)$, 
for the $D^{* \pm}$ candidates for the LER ($\sqrt{s}=225$\,GeV) data sample.  Other 
details as in Fig.~\ref{fig:dstar-her}.}
\label{fig:dstar-ler}
\end{figure}

The distributions of $\Delta M$ for $D^{*+}$ candidates in the HER, MER and LER periods, 
without the requirement on $\Delta M$, are shown in Figs.~\ref{fig:dstar-her}--\ref{fig:dstar-ler}.  
Clear $D^{*+}$ peaks are seen.  The $D^{*+}$ signal was extracted by subtracting the 
correct-sign background estimate from the number of candidates in the signal window 
$0.143<\Delta M<0.148$\,GeV.  The shape of the background was determined by performing 
a simultaneous fit to the correct-sign and wrong-sign distributions, as outlined in 
a previous publication~\cite{jhep:05:097}.  
The fit was performed in the region $\Delta M < 0.168$\,GeV; the region with a possible 
signal contribution, $0.140 < \Delta M < 0.150$\,GeV, was removed from the fit to the 
correct-sign distribution.  The total signals are 
$N_{\rm HER}^{D^*} = 12256 \pm 191$, $N_{\rm MER}^{D^*} = 417 \pm 37$ and 
$N_{\rm LER}^{D^*} = 859 \pm 49$ for the HER, MER and LER samples, respectively. 


\section{Monte Carlo samples}
\label{sec:mc}

The acceptance and effects of detector response were determined using samples 
of simulated events.  The Monte Carlo (MC) programme 
{\sc Pythia}~6.221~\cite{cpc:82:74,*cpc:135:238}, which implements leading-order 
matrix elements, followed by parton showers and hadronisation, was used.  
Different subprocesses were generated separately~\cite{epj:c71:1659}. 
The CTEQ5L~\cite{pr:d55:1280} and GRV-LO~\cite{pr:d45:3986,*pr:d46:1973} sets were 
used for the proton and photon parton density functions (PDFs), respectively.  Samples of charm and 
beauty events were generated with quark masses, $m_c=1.5$\,GeV and $m_b=4.75$\,GeV.

The generated MC events were passed through the ZEUS detector and trigger simulation 
programmes based on {\sc Geant} 3.21~\cite{tech:cern-dd-ee-84-1}.  They were then 
reconstructed and analysed using the same programmes as used for the data.


\section{QCD calculations}
\label{sec:qcd}

The data are compared with an NLO QCD prediction from Frixione et 
al.~\cite{pl:b348:633,*np:b454:3} in the fixed-flavour-number scheme (FFNS), in which only light 
flavours and gluons are present as partons in the proton and heavy quarks are produced 
in the hard interaction~\cite{np:b374:36}.  The following input parameters were set in 
the calculation: the renormalisation and factorisation scales were set to 
$\mu= \sqrt{m_c^2 + \hat{p}_T^2}$, where $\hat{p}_T$ is the average transverse 
momentum of the charm quarks and the pole mass was $m_c=1.5$\,GeV; the 
proton and photon PDFs were ZEUS-S 3-flavour FFNS~\cite{pr:d67:012007} 
and GRV-G~HO~\cite{pr:d45:3986,*pr:d46:1973}; the value of the strong coupling constant 
was $\alpha_s(M_Z) = 0.118$ for five flavours;
and the parameter, $\epsilon$, in the Peterson fragmentation function~\cite{pr:d27:105} 
was $\epsilon = 0.079$~\cite{jhep:04:082}.  The contribution to the $D^{*+}$ visible 
cross section from beauty production is predicted by MC to be about 2.5\%.
This value was the same to within 0.1\% for all three data sets.  Therefore, the beauty
contribution cancelled and the uncertainty was negligible when the cross sections were 
normalised.  Hence, the beauty component was not included in the predictions.

Several sources of theoretical uncertainty were investigated and are listed 
in the following, with the respective effects on the (MER, LER) samples normalised to the HER data 
given in parentheses:

\begin{itemize}

\item the renormalisation and factorisation scales were changed independently to 0.5 
and 2 times their nominal value.  The largest change in the positive and negative 
direction was taken as the systematic uncertainty $(^{+3.5}_{-1.6}\%$, $^{+5.2}_{-2.3}\%)$;

\item the fragmentation parameter $\epsilon$ was varied in the 
range~\cite{jhep:04:082,epj:c59:589} from 0.006 to 0.092 ($^{+1.5}_{-0.1}\%$, $^{+2.3}_{-0.2}\%$);

\item the proton PDF was changed to the ABM11 3-flavour FFNS~\cite{pl:b699:345} 
      parametrisations ($+0.9\%$, $+1.3\%$);

\item the value of $m_c$ was changed to 1.35 and 1.65\,GeV ($^{+0.1}_{-0.2}\%$, $^{+0.1}_{-0.3}\%$).

\end{itemize}


\section{Determination of normalised cross sections}
\label{sec:xsec-method}

\begin{figure}[p]
\begin{center}
\epsfig{file=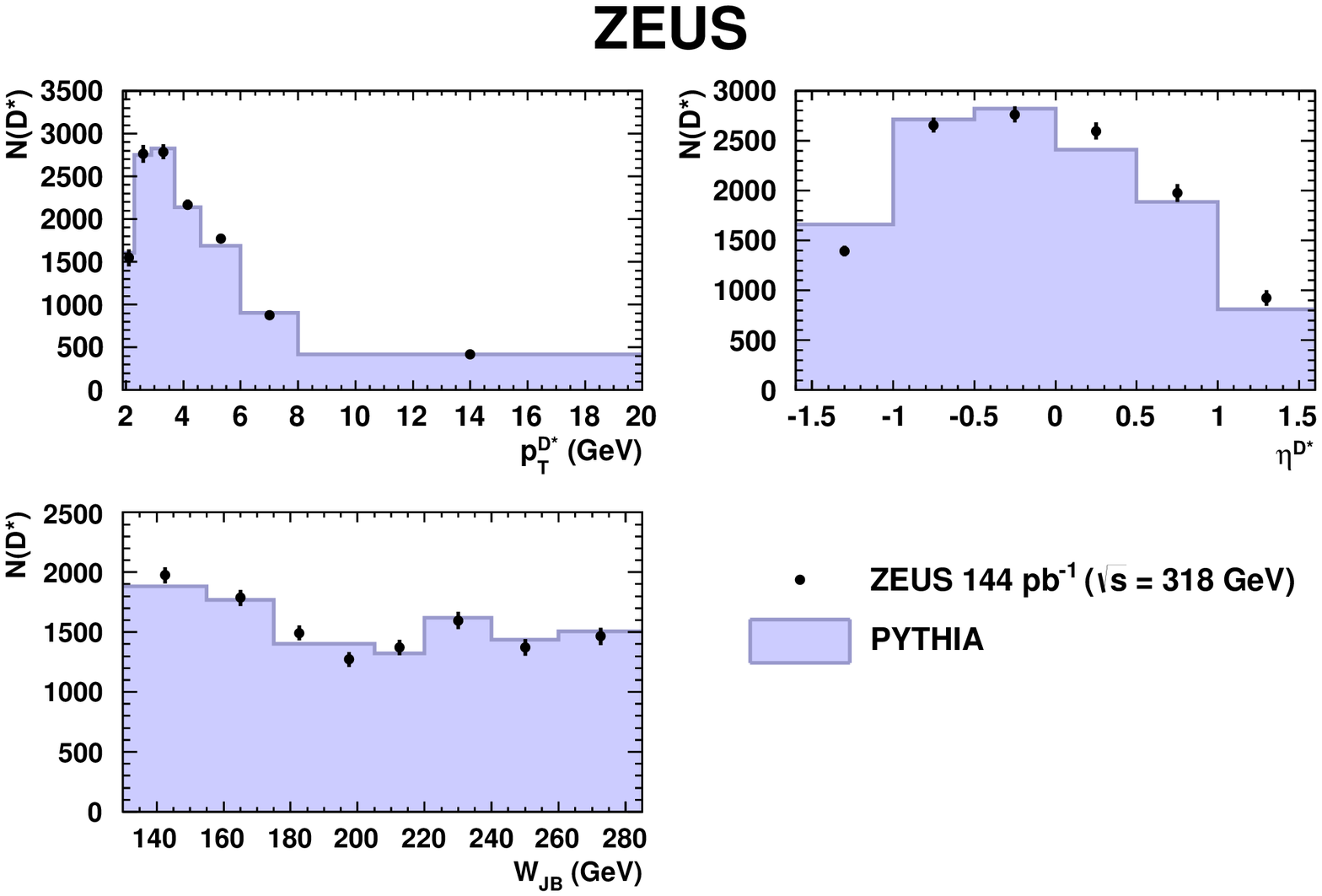,width=\textwidth}
\put(-280,267){\makebox(0,0)[tl]{\large (a)}}
\put(-55,267){\makebox(0,0)[tl]{\large (b)}}
\put(-280,123){\makebox(0,0)[tl]{\large (c)}}
\end{center}
\caption{Distributions of (a) $p_T^{D^*}$, (b) $\eta^{D^*}$ and (c) $W_{\rm JB}$ 
for $D^*$ mesons in the HER ($\sqrt{s} =318$\,GeV) data sample (points) compared 
with a mixture of charm and beauty events from the {\sc Pythia} MC simulation 
(histogram).
}
\label{fig:control-HER}
\end{figure}

\begin{figure}[p]
\begin{center}
\epsfig{file=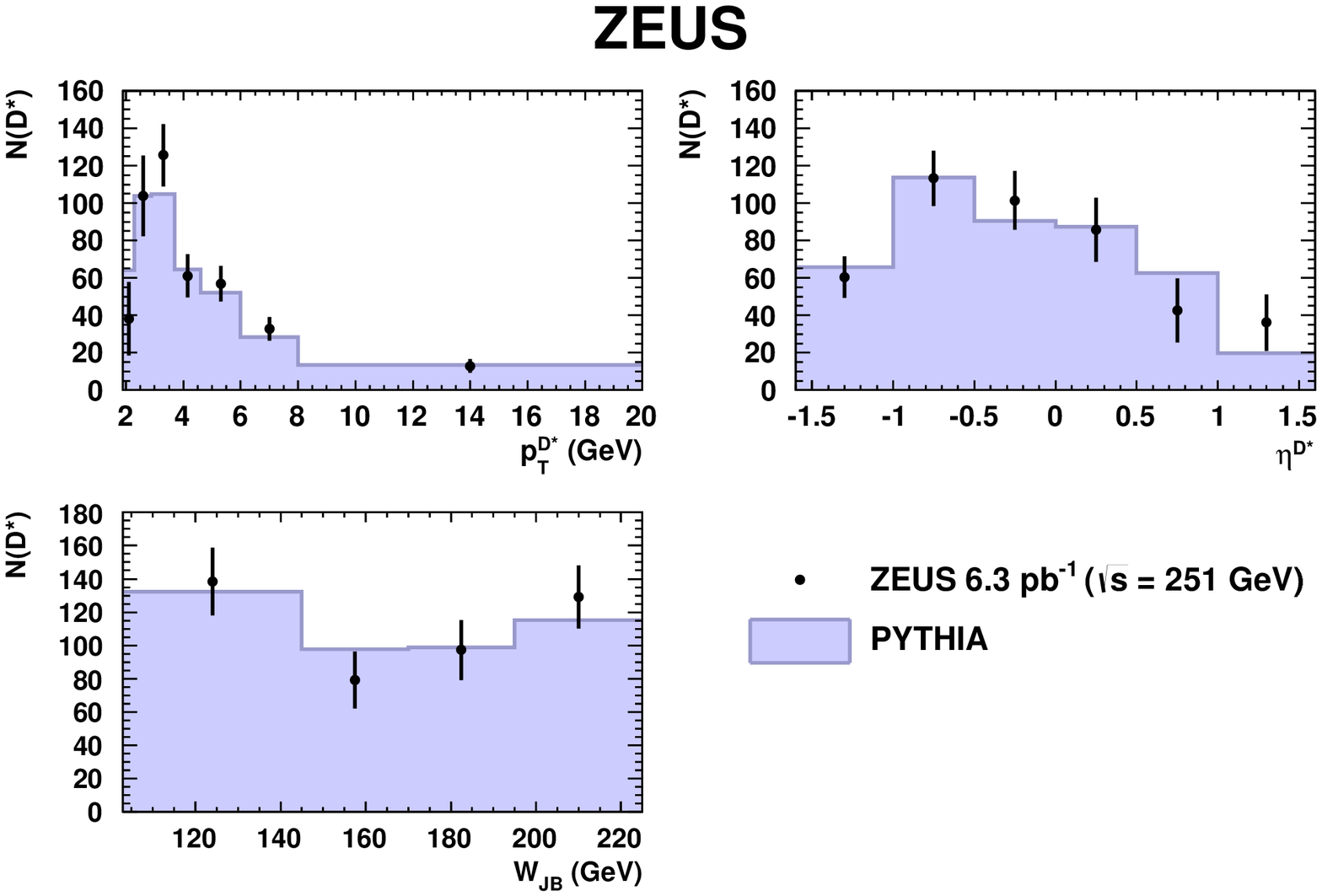,width=\textwidth}
\put(-280,267){\makebox(0,0)[tl]{\large (a)}}
\put(-55,267){\makebox(0,0)[tl]{\large (b)}}
\put(-280,123){\makebox(0,0)[tl]{\large (c)}}
\end{center}
\caption{Distributions of (a) $p_T^{D^*}$, (b) $\eta^{D^*}$ and (c) $W_{\rm JB}$ 
for $D^*$ mesons in the MER ($\sqrt{s} =251$\,GeV) data sample (points) compared 
with a mixture of charm and beauty events from the {\sc Pythia} MC simulation 
(histogram).
}
\label{fig:control-MER}
\end{figure}

\begin{figure}[p]
\begin{center}
\epsfig{file=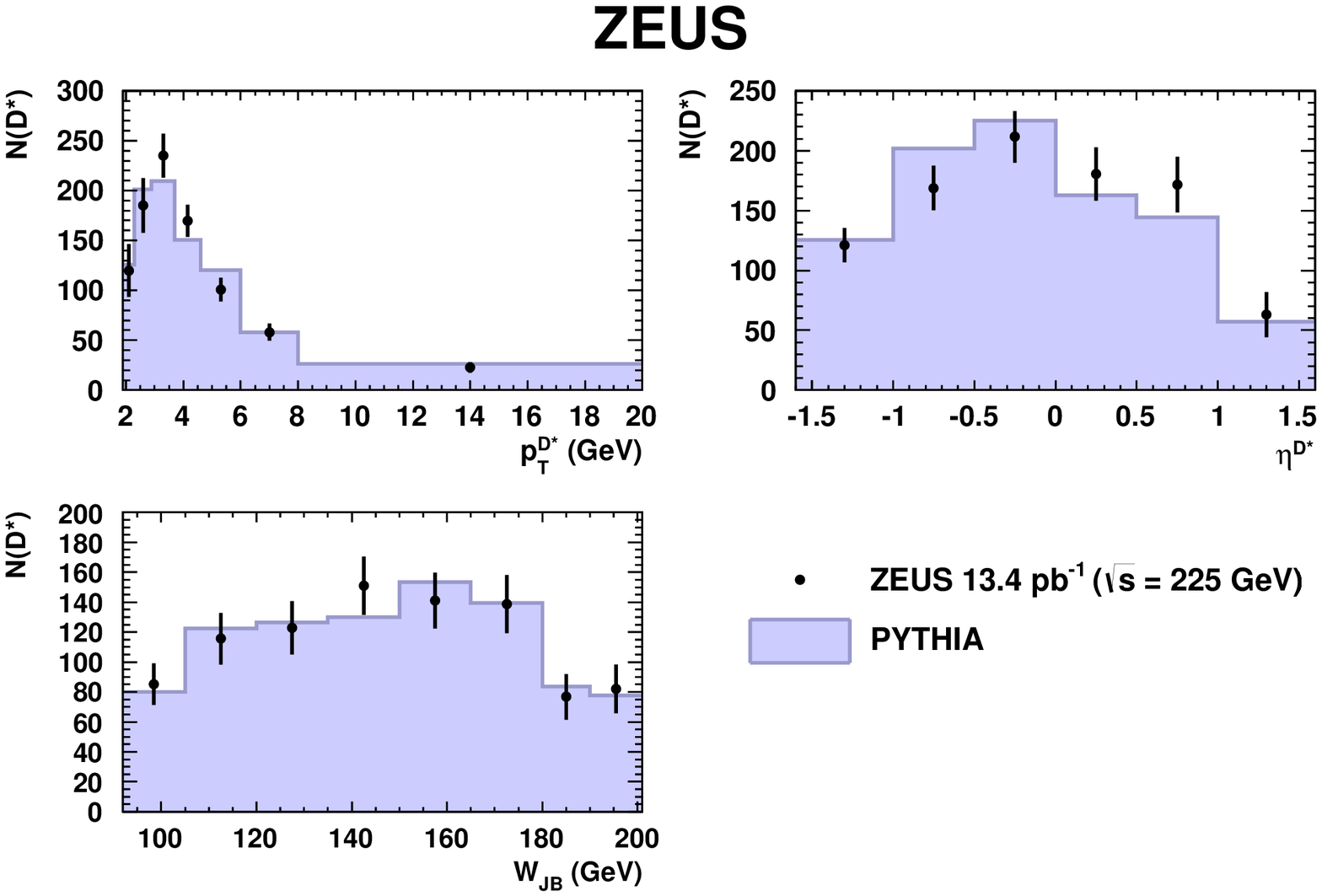,width=\textwidth}
\put(-280,267){\makebox(0,0)[tl]{\large (a)}}
\put(-55,267){\makebox(0,0)[tl]{\large (b)}}
\put(-280,123){\makebox(0,0)[tl]{\large (c)}}
\end{center}
\caption{Distributions of (a) $p_T^{D^*}$, (b) $\eta^{D^*}$ and (c) $W_{\rm JB}$ 
for $D^*$ mesons in the LER ($\sqrt{s} =225$\,GeV) data sample (points) compared 
with a mixture of charm and beauty events from the {\sc Pythia} MC simulation 
(histogram).
}
\label{fig:control-LER}
\end{figure}

Visible $D^{*+}$ photoproduction cross sections in the kinematic region 
$1.9 < p_T^{D^*} < 20$\,GeV, $|\eta^{D^*}|<1.6$, $Q^2<1$\,GeV$^2$ and 
$0.167 < y < 0.802$ were obtained using the formula

\[
\sigma_{\rm vis} = \frac{N_{\rm data}^{D^*}}{{\mathcal A} \cdot BR \cdot {\mathcal L}}\,,
\]

where $N_{\rm data}^{D^*}$ is the number of $D^{*+}$ mesons in the data, $BR$ is the product 
of the branching fractions of the decay $D^{*+} \to D^0 \pi_s^+$ 
with $D^0 \to K^- \pi^+$ and ${\mathcal L}$ 
is the integrated luminosity of the respective sample.  The acceptance, ${\mathcal A}$, 
is given by the ratio of the number of reconstructed to generated $D^{*+}$ mesons in the 
MC simulation, using a mix of charm and beauty production.  The sample of beauty MC 
events, both reconstructed and generated, was scaled by a factor of 1.6, 
consistent with previous ZEUS measurements~\cite{epj:c71:1659,pr:d78:072001,jhep:04:133}.  
In order to optimise the description of the data and hence determine the acceptances as 
accurately as possible, the MC was reweighted in 
$W_{\rm JB}$ for the HER sample and in $p_T^{D^*}$ for the HER, MER and LER data samples.  
The comparison of background-subtracted data and MC after these reweightings is shown 
in Figs.~\ref{fig:control-HER},~\ref{fig:control-MER} and~\ref{fig:control-LER}, 
for the HER, MER and LER samples, respectively.  The description of the data is reasonable, 
also for the $\eta^{D^*}$ distributions, for which no reweighting was performed.  

The measured cross sections were normalised to the HER data sample:

\[
R^{\rm HER,MER,LER}_{\sigma} = \sigma^{\rm HER,MER,LER}_{\rm vis}/\sigma^{\rm HER}_{\rm vis}.
\]
  
This allowed the energy 
dependence of the cross section to be studied to higher precision as a number 
of systematic uncertainties in data and theory cancel.

The following sources of systematic uncertainty were considered~\cite{thesis:zakharchuk:2014}, 
with the effect on $R^{\rm MER}_{\sigma}$ and $R^{\rm LER}_{\sigma}$ given in parentheses:

\begin{itemize}

\item the lower and upper $W_{\rm JB}$ cuts for data and reconstructed MC events were 
changed by $\pm 5$\,GeV in order to assess the effects of the resolution of $W_{\rm JB}$ 
and the impact of any residual backgrounds ($^{+0.7}_{-0.8}\%$, $^{+2.1}_{-2.1}\%$);

\item the forms of the functions used for MC reweighting in $W_{\rm JB}$ (HER only) and 
$p_T^{D^*}$ were varied within the uncertainties determined from the quality of the 
description of the data ($^{+1.4}_{-1.4}\%$, $^{+3.2}_{-1.3}\%$);

\item the lower and upper mass requirements for the $D^0$ were varied to 1.80\,GeV and 
1.93\,GeV, both in data and MC.  This and the following two 
sources were performed to assess the uncertainty coming from estimation of the combinatorial 
background ($-6.7\%$, $^{+0.7}_{-4.1}\%$);

\item the upper edge of the fit range in the $\Delta M$ distribution was changed to 
0.165\,GeV, both in data and MC ($-0.7\%$, $-1.9\%$);

\item the minimum requirement on the ratio $p_T^{D^{*}}/E_T^{\theta>10^\circ}$ 
was varied between 0.05 and 0.20, both in data and MC ($^{+2.0}_{-2.3}\%$, $^{+2.1}_{-1.1}\%$);

\item the uncorrelated uncertainty in the luminosity determination ($\pm 1.4\%$, $\pm 1.4\%$).

\end{itemize}

The above systematic uncertainties were added in quadrature separately for positive and 
negative variations.
Other sources of systematic uncertainty were found to be negligible and were ignored.  
These included the uncertainties on the track-finding efficiency, additional reweighting of the 
MC samples in $\eta^{D^*}$ as well as from the fraction 
of beauty events used in the acceptance correction.  As a cross-check, 
the number of $D^{*+}$ mesons was also extracted by subtracting the wrong-sign from the 
correct-sign distribution; the result was consistent with the nominal procedure.

The statistical uncertainties for $R^{\rm MER}_{\sigma}$ and $R^{\rm LER}_{\sigma}$ include 
that from the HER sample, although the uncertainties from the MER and LER dominate.  
The systematic uncertainties also contain contributions from the HER result which are 
fully correlated between the LER and MER measurements.  


\section{Energy dependence of \boldmath{$D^{*+}$} cross sections}
\label{sec:xsec-results}

Ratios of visible $D^{*+}$ photoproduction cross sections have been measured in the 
kinematic region $1.9 < p_T^{D^*} < 20$\,GeV, $|\eta^{D^*}|<1.6$, $Q^2<1$\,GeV$^2$ 
and $0.167 < y < 0.802$.  The range in $y$ corresponds to photon--proton centre-of-mass 
energy, $W$, ranges of 
$130 < W < 285$\,GeV, $103 < W < 225$\,GeV and 
$92 < W < 201$\,GeV in the HER, MER, LER samples, respectively.  
The ratios of the visible cross sections for the MER and LER samples to that of the HER 
sample are:

\vspace{-1cm}
\begin{eqnarray}
R_{\sigma}^{\rm MER} & = & 0.780 \pm 0.074 ({\rm stat.}) ^{+0.022}_{-0.058} ({\rm syst.}) \nonumber \\
R_{\sigma}^{\rm LER} & = & 0.786 \pm 0.049 ({\rm stat.}) ^{+0.037}_{-0.043} ({\rm syst.})\,. \nonumber 
\end{eqnarray}

\begin{figure}[p]
\begin{center}
\epsfig{file=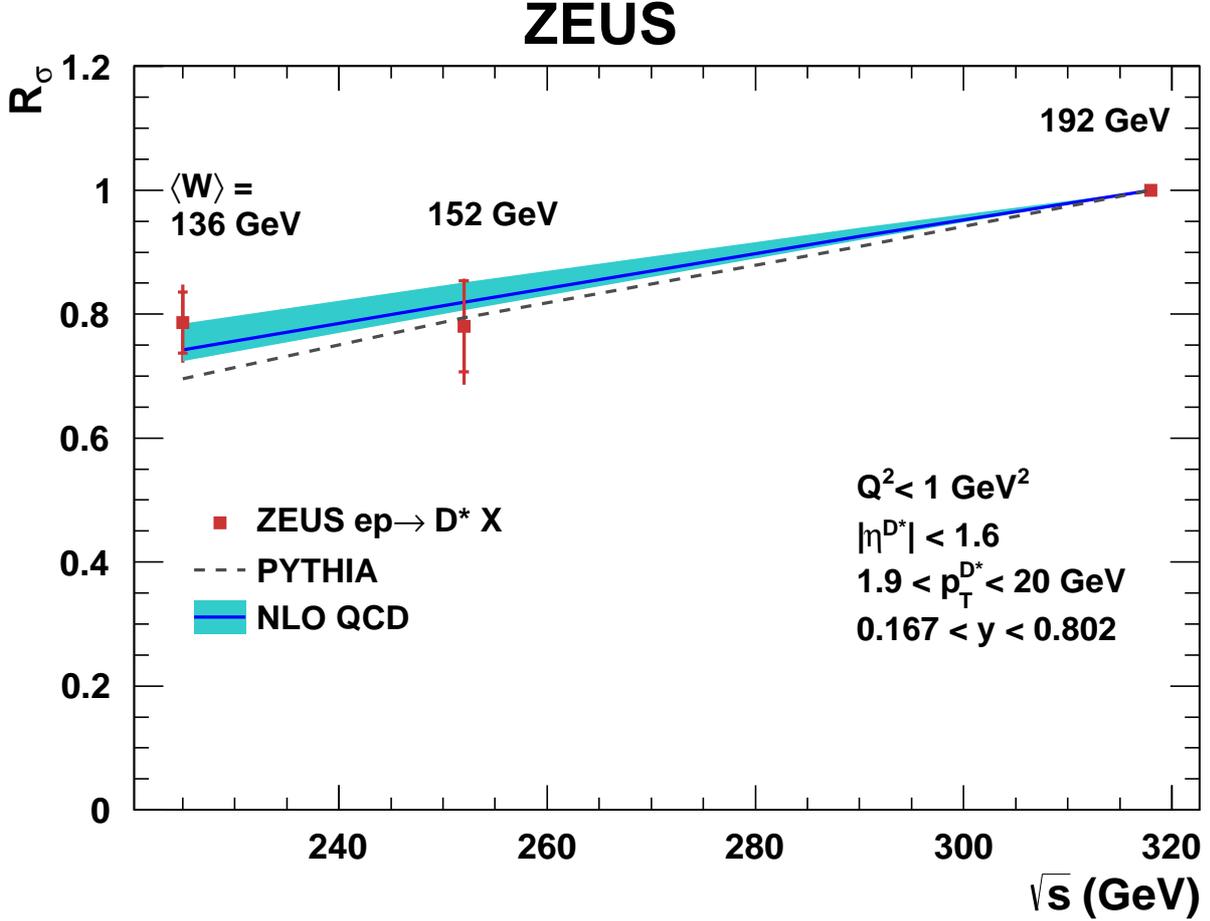,width=\textwidth}
\end{center}
\caption{
Normalised $D^*$ visible photoproduction cross sections as a function of the 
$ep$ centre-of-mass energy.  The data (points) are shown with statistical 
uncertainties (inner error bars) and statistical and systematic uncertainties 
added in quadrature (outer error bars).  The predictions from NLO QCD (solid line) 
are shown with the uncertainties given in Section~\ref{sec:qcd} added in quadrature 
separately for positive and negative variations (band).  A prediction from the {\sc Pythia} 
MC simulation is also shown (dashed line).  The data and theory 
at $\sqrt{s}=318$\,GeV are constrained by definition to be at unity, with no uncertainty.  
At each data point, the average photon--proton centre-of-mass energy, $\langle W \rangle$, 
is also given.   
}
\label{fig:result}
\end{figure}

These values, along with $R_{\sigma}^{\rm HER}$ (by constraint equal to unity), 
are shown in Fig.~\ref{fig:result}.  The cross sections for the MER and LER samples are 
compatible within uncertainties, but significantly smaller than the cross section for the 
HER data.  This behaviour of 
increasing cross section with increasing $ep$ centre-of-mass energy is predicted well by 
{\sc Pythia} MC simulations and NLO QCD, although the predictions have a somewhat different slope.  This shows that the proton PDFs constrained primarily from inclusive deep inelastic 
scattering data are able to describe this complementary process which probes in particular 
the gluon distribution.  
The physics possibilities of future colliders such as the Large Hadron Electron Collider 
(LHeC)~\cite{natp:9:448,*jp:g39:075001} are studied using current NLO QCD calculations.  
The results shown here enhance confidence in the NLO QCD predictions 
of charm production rates, specifically, and QCD processes, in general, for a future 
TeV-scale $ep$ collider.


\section{Summary}
\label{sec:summary}

Photoproduction of $D^{*\pm}$ mesons has been measured at HERA at three different $ep$ 
centre-of-mass energies, $\sqrt{s} = 318$, 251 and 225\,GeV.  For $D^{*\pm}$ mesons in the 
range $1.9 < p_T^{D^*} < 20$\,GeV and $|\eta^{D^*}|<1.6$, cross sections normalised to the 
result at $\sqrt{s} = 318$\,GeV were measured.  Photoproduction events were selected in the range 
$Q^2 < 1$\,GeV$^2$ and $0.167 < y < 0.802$ where the range in $y$ corresponds to the 
photon--proton centre-of-mass energies of $130 < W < 285$\,GeV, 
$103 < W < 225$\,GeV and $92 < W < 201$\,GeV.  The cross sections, 
normalised to that for the highest $\sqrt{s}$, show an increase with increasing $\sqrt{s}$.  
This is predicted well by perturbative QCD, demonstrating consistency of the gluon distribution 
probed here with that extracted in PDF fits to inclusive deep inelastic scattering data.

\clearpage

\section*{Acknowledgements}
\label{sec-ack}

\Zacknowledge

\clearpage

{
\ifzeusbst
  \ifzmcite
     \bibliographystyle{./BiBTeX/bst/l4z_default3}
  \else
     \bibliographystyle{./BiBTeX/bst/l4z_default3_nomcite}
  \fi
\fi
\ifzdrftbst
  \ifzmcite
    \bibliographystyle{./BiBTeX/bst/l4z_draft3}
  \else
    \bibliographystyle{./BiBTeX/bst/l4z_draft3_nomcite}
  \fi
\fi
\ifzbstepj
  \ifzmcite
    \bibliographystyle{./BiBTeX/bst/l4z_epj3}
  \else
    \bibliographystyle{./BiBTeX/bst/l4z_epj3_nomcite}
  \fi
\fi
\ifzbstjhep
  \ifzmcite
    \bibliographystyle{./BiBTeX/bst/l4z_jhep3}
  \else
    \bibliographystyle{./BiBTeX/bst/l4z_jhep3_nomcite}
  \fi
\fi
\ifzbstnp
  \ifzmcite
    \bibliographystyle{./BiBTeX/bst/l4z_np3}
  \else
    \bibliographystyle{./BiBTeX/bst/l4z_np3_nomcite}
  \fi
\fi
\ifzbstpl
  \ifzmcite
    \bibliographystyle{./BiBTeX/bst/l4z_pl3}
  \else
    \bibliographystyle{./BiBTeX/bst/l4z_pl3_nomcite}
  \fi
\fi
{\raggedright
\bibliography{./syn.bib,%
              ./myref.bib,%
              ./BiBTeX/bib/l4z_zeus.bib,%
              ./BiBTeX/bib/l4z_h1.bib,%
              ./BiBTeX/bib/l4z_articles.bib,%
              ./BiBTeX/bib/l4z_books.bib,%
              ./BiBTeX/bib/l4z_conferences.bib,%
              ./BiBTeX/bib/l4z_misc.bib,%
              ./BiBTeX/bib/l4z_preprints.bib}}
}
\vfill\eject

%
%
\end{document}